\documentclass[aps,prb,twocolumn,superscriptaddress]{revtex4}

\usepackage{graphicx,color}

\input epsf.sty
\usepackage{graphicx}

\begin{document}

\preprint{\today}

%
%

\title{Inelastic neutron scattering study on the resonance mode in an optimally doped superconductor LaFeAsO$_{0.92}$F$_{0.08}$}

\author{Shin-ichi Shamoto}
\affiliation{ Quantum Beam Science Directorate, Japan Atomic Energy Agency,
   Tokai, Ibaraki 319-1195, Japan }
\affiliation{ JST, Transformative Research-Project on Iron Pnictides (TRIP),
   Tokyo 102-0075, Japan }

\author{Motoyuki Ishikado}
\affiliation{ Quantum Beam Science Directorate, Japan Atomic Energy Agency,
   Tokai, Ibaraki 319-1195, Japan }
\affiliation{ JST, Transformative Research-Project on Iron Pnictides (TRIP),
   Tokyo 102-0075, Japan }

\author{Andrew D. Christianson}
\affiliation{ Oak Ridge National Laboratory, Oak Ridge, TN 37831, USA }

\author{Mark D. Lumsden}
\affiliation{ Oak Ridge National Laboratory, Oak Ridge, TN 37831, USA }

\author{Shuichi Wakimoto}
\affiliation{ Quantum Beam Science Directorate, Japan Atomic Energy Agency,
   Tokai, Ibaraki 319-1195, Japan }
\affiliation{ JST, Transformative Research-Project on Iron Pnictides (TRIP),
   Tokyo 102-0075, Japan }

\author{Katsuaki Kodama}
\affiliation{ Quantum Beam Science Directorate, Japan Atomic Energy Agency,
   Tokai, Ibaraki 319-1195, Japan }
\affiliation{ JST, Transformative Research-Project on Iron Pnictides (TRIP),
   Tokyo 102-0075, Japan }

\author{Akira Iyo}
\affiliation{ Nanoelectronics Research Institute, National Institute of Advanced
   Industrial Science and Technology, Tsukuba, Ibaraki 305-8562, Japan }
\affiliation{ JST, Transformative Research-Project on Iron Pnictides (TRIP),
   Tokyo 102-0075, Japan }

\author{Masatoshi Arai}
\affiliation{ J-PARC Center, Japan Atomic Energy Agency,
   Tokai, Ibaraki 319-1195, Japan }
\affiliation{ JST, Transformative Research-Project on Iron Pnictides (TRIP),
   Tokyo 102-0075, Japan }

\date{\today}

\begin{abstract}

An optimally doped iron-based superconductor LaFeAsO$_{0.92}$F$_{0.08}$ with $T_c = 29$~K has been studied by inelastic powder neutron scattering.  The magnetic excitation at $Q=1.15$~\AA$^{-1}$ is enhanced below $T_c$, leading to a peak at $E_{res}\sim13$~meV as the resonance mode, in addition to the formation of a gap at low energy below the crossover energy $\Delta_{c}\sim10 meV$. The peak energy at $Q=1.15$~\AA$^{-1}$ corresponds to $5.2 k_B T_c$ in good agreement with the other values of resonance mode observed in the various iron-based superconductors.  Although the phonon density of states has a peak at the same energy as the resonance mode in the present superconductor, the $Q$-dependence is consistent with the resonance being of predominately magnetic origin.

\end{abstract}


\maketitle

Extensive research on iron based superconductors was initiated by the discovery of superconductivity in LaFeAsO$_{1-x}$F$_{x}$~\cite{Kamihara_08}.  Higher superconducting transition temperatures, $T_c$, up to 55~K have been discovered in the $Ln$FeAsO$_{1-x}$F$_{x}$ family of materials by replacing La with rare earth elements with smaller ionic radii \cite{Ren_08,Kito_08,Chen_08}.  Although many superconductors with iron pnictide/chalcogenide layers such as (Ba,K)Fe$_{2}$As$_{2}$~\cite{Rotter_08}, LiFeAs~\cite{Wang_08}, and FeSe~\cite{Hsu_08} have been discovered, the highest $T_c$ in iron-based superconductors is achieved in the $Ln$FeAsO$_{1-x}$F$_{x}$ materials.  Therefore, a comprehensive understanding of basic physical properties, such as the magnetic excitation spectrum in the $Ln$FeAsO$_{1-x}$F$_{x}$ superconductors is essential for the progress in this field.

In the superconducting state, a resonance in the spin excitation spectrum has been observed by inelastic neutron scattering in several iron-based superconductors, such as, Ba$_{0.6}$K$_{0.4}$Fe$_{2}$As$_{2}$~\cite{Christianson_08}, Ba(Fe$_{1-x}$Co$_{x}$)$_{2}$As$_{2}$~\cite{Lumsden_09,Christianson_09,Inosov_09}, BaFe$_{1.9}$Ni$_{0.1}$As$_{2}$~\cite{Chi_09}, FeTe$_{1-x}$Se$_{x}$~\cite{Mook_09,Qiu_09}. The resonance is characterized by an increase of magnetic scattering intensity below $T_c$ and has been considered as a finger print of a sign-change of the superconducting gap function between the cylindrical $\Gamma$ and $M$-point Fermi surfaces (FS).  An extended $s_{\pm}$ pairing is a candidate consistent with this observation \cite{Maier_08,Korshunov_08}. However, a resonance has not yet been observed in the $Ln$FeAsO$_{1-x}$F$_{x}$ system due to the lack of availability of large single crystals~\cite{Ishikado_09}.  Therefore, we have extensively studied the LaFeAsO$_{1-x}$F$_{x}$ system by inelastic powder neutron scattering~\cite{Ishikado_JPSJ09,Waki_09}.  $Ln$=La has been selected since it contains no 4f electrons and, hence, there are no crystal field excitations to complicate the excitation spectrum.  We note that the initial observation of the resonance was carried out on a polycrystalline sample of Ba$_{0.6}$K$_{0.4}$Fe$_{2}$As$_{2}$~\cite{Christianson_08}. Here, we report the temperature, energy, and momentum dependence of the magnetic resonance mode in optimally doped LaFeAsO$_{0.92}$F$_{0.08}$ with $T_c=29$~K.

A polycrystalline sample of LaFeAsO$_{0.92}$F$_{0.08}$ was prepared by solid-state reaction using Fe$_{2}$O$_{3}$, Fe, LaAs, and FeF$_{2}$ powders as starting materials.  LaFe was obtained by reacting La powders and As pieces at 500~$^{\circ}$C for 10~h and then heating at 850~$^{\circ}$C for 5~h.  (The starting materials were mixed at nominal compositions of LaFeAs(O,F)$_{0.9}$ because of the partial oxidation of the precursor LaAs.)  They were then ground using an agate mortar in a glove box filled with dry nitrogen gas.  The raw materials were thoroughly mixed and pressed into pellets.  The pellets were wrapped with Ta foil and sealed in an evacuated quartz tube.  They were then annealed at 1100~$^{\circ}$C for 10~h.  Powder X-ray diffraction (XRD) analysis showed no impurity phase.  The XRD pattern is well indexed on the basis of the tetragonal structure with the space group $P4/nmm$ ($a=4.026$~\AA~ and $c=8.723$~\AA).  The fluorine content was determined by the secondary ion-microprobe mass spectroscopy.  Superconductivity of the prepared sample was characterized by SQUID measurement in a cooling process under a magnetic field of 5~Oe, leading to about 40~\% volume fraction with $T_c$ = 29~K.

We performed inelastic neutron scattering measurements using the HB-3 spectrometer at the HFIR facility in the Oak Ridge National Laboratory, USA on a 25~g polycrystalline sample.  Collimations of 48$'$-80$'$-S-80$'$-240$'$ (S denotes sample) and fixed final energies of $E_f=30.5$~meV and 14.7~meV were utilized, leading to instrumental resolutions of 4.0 and 1.4~meV at the elastic position and wavevector resolutions of 0.057 and 0.042~\AA$^{-1}$, respectively.

Constant energy scans at $T=35$ K are presented in Fig. 1 (a) and show the magnetic excitations as broad peaks centered at $Q=1.15$~\AA$^{-1}$.  The wavevector is the same as that of the magnetic rod observed in the parent compound LaFeAsO above the spin density wave transition temperature $T_{~\rm AF}$~\cite{Ishikado_JPSJ09}.  The reciprocal lattice position roughly corresponds to (1/2 1/2 0) in the tetragonal unit cell, which is associated with FS nesting vectors from the hole FS cylinders along the $\Gamma-Z$ line to electron FS cylinders along the $M-A$ line in the first Brillouin zone (BZ)~\cite{Singh_08}.  As the scattering is expected to be two-dimensional, inelastic neutron scattering will measure the powder average of four magnetic rods centered at \{1/2 1/2 0\}.   This averaging will lead to a slight peak shift from $Q=1.10$~\AA$^{-1}$ at (1/2 1/2 0) to $Q=1.15$~\AA$^{-1}$, resulting in a weak tail on the large $Q$ side~\cite{Ishikado_JPSJ09}.  We define this peak position as $Q_{\rm AF}$.  The peak asymmetry becomes obscure even in the parent compound at high temperature due to the short correlation length~\cite{Ishikado_JPSJ09}.  Therefore, the $Q$-dependence of the peak can be approximated by a Gaussian function.  The peak width at $T=35$~K in Fig. 1 (a) increases with increasing energy transfer.  Assuming that the $E$-dependence of the width originates from an in-plane antiferromagnetic spin wave dispersion, the present in-plane spin wave band width is roughly 50~meV.  This value is comparable to that observed in BaFe$_{1.84}$Co$_{0.16}$As$_{2}$~\cite{Lumsden_09}, where the in-plane spin wave bandwidth was estimated to be 70 meV.  Below $T_c$ the $Q$-width near the expected resonance energy becomes sharper.  This can be attributed to the additional scattering of the spin resonance.

 As for the absolute scale in the present measurement, the total $Q$-integrated dynamical spin susceptibility at $E$=11 meV was about $4$~$\mu_B$/eV/Fe based on (0 0 2) integrated Bragg peak intensity.  This result is consistent with the other measurements at JRR-3 in Japan~\cite{Waki_09}.  We also note that the neutron scattering measurements on the present sample show no evidence of long range magnetic order at $T=4$~K.

\begin{figure}
\includegraphics[width=3.0in]{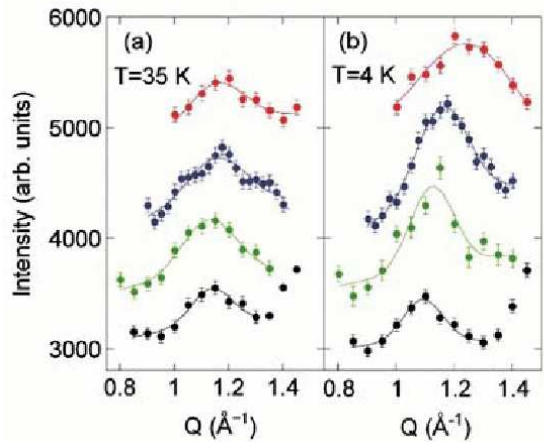}
\caption{(Color online)
(a) Constant-$E$ scans with a fixed final energy of 30.5 meV at $T=35$ K as a function of $Q$ for $E$=9 (black), 11 (green), 13 (blue) and 15 (red) meV in order from the bottom to the top. (b) Constant-$E$ scans at $T=4$ K as a function of $Q$ for $E$=9 (black), 11 (green), 13 (blue) and 15 (red) meV. 1000 has been added to the intensity at $E=15$ meV, while at $E=11$ meV 300 has been subtracted for clarity.
}
\end{figure}

The energy dependence of the dynamical structure factor $S(Q_{\rm AF}, E)$ at $T=4$ and 35~K is shown in Fig. 2.  The background (BG) intensity at $Q_{\rm AF}$=1.15~$\AA^{-1}$ has been estimated from the constant-E scans in Fig. 1 and has a peak at $E=13$~meV.  This peak is consistent with the the phonon density of states of LaFeAsO$_{1-x}$F$_{x}$ ~\cite{Christianson_PRL08, Fukuda_08}.  At $T=35$~K, most of the observed scattering comes from the non-magnetic background while at $T=4$~K additional intensity centered at 13~meV appears.  We define the peak energy at $Q_{AF}$ as $E_{res}$. The peak is shown clearly in the difference plot of Fig. 2 (c). The upturn in scattering at low energies (Fig. 2 (a) and (b)) is incoherent elastic scattering, being consistent with the energy resolution with final neutron energies of 30.5 and 14.7 meV.  Considerable spectral weight (about 10~\% assuming a 50~meV bandwidth) is shifted from low energies to the resonance peak, where the crossover energy, $\Delta_c$, is about 10 meV. On the other hand, the total $S(Q_{\rm AF}, E)$ weight is roughly preserved.

The temperature dependence of $S(Q_{\rm AF}, E)$ at $E$=11 and 13 meV is shown in Fig. 3.  There appears to be a strong correlation between the magnetic excitation and the superconductivity at two energies above $\Delta_c$.  This supports the contention that this magnetic excitation is the resonance mode as predicted~\cite{Maier_08,Korshunov_08}.  The peak intensity at $E=11$~meV saturates rapidly while that at $E=13$~meV grows slowly with decreasing temperature.  At $E=11$~meV, the intensity saturation at low temperature may be due to the coarser energy resolution with $E_f=30.5$~meV, which covers both of negative and positive regions in Fig. 2 (c).  The temperature dependence at 13 meV is consistent with the temperature dependence seen in other Fe-based superconductors \cite{Christianson_08,Lumsden_09,Christianson_09}. These temperature dependences of the resonance peak intensity may be attributed to the temperature dependence of $\Delta_c$~\cite{Inosov_09}.

\begin{figure}
\includegraphics[width=3in]{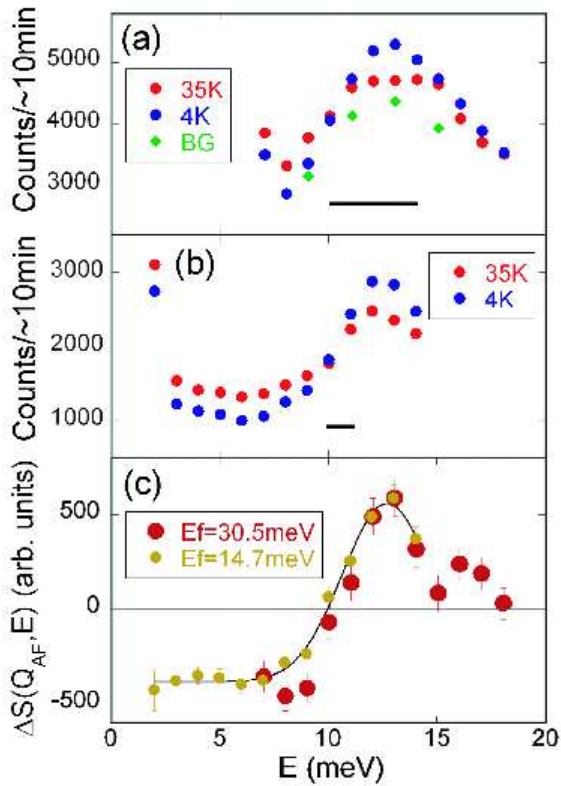}
\caption{(Color online)
(a) Constant-$Q$ scans at $T=4$ and 35 K as a function of $E$ for $Q=1.15$ \AA$^{-1}$ at $E_{f}=30.5$~meV.  BG is estimated background from the fitting at each energy in Fig. 1 (b). (b) Constant-$Q$ scans at $T=4$ and 35 K as a function of $E$ for $Q=1.15$ \AA$^{-1}$ at $E_{f}=14.7$ meV. (c) The $4-35$ K difference of dynamical structure factor $S(Q_{\rm AF}, E)$ for constant-$Q$ scans measured at $Q=1.15$ \AA$^{-1}$. $\Delta S(Q_{\rm AF}, E)$ of $E_{f}=14.7$ meV is enlarged to agree with that at 13 meV of $E_{f}=30.5$ meV. Both results are consistent with each other, suggesting the broad feature in energy relative to their $E$-resolutions. Horizontal bars are the $E$-resolutions. The solid line is a guide to the eye.
}
\end{figure}

\begin{figure}
\includegraphics[width=2.5in]{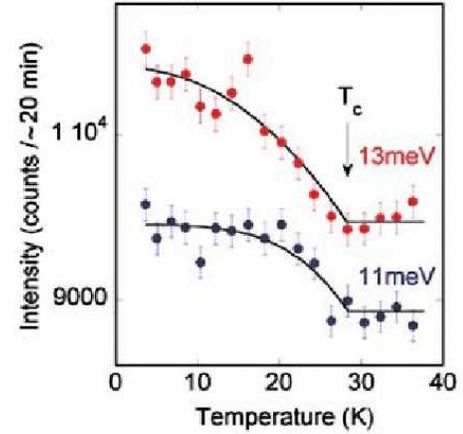}
\caption{(Color online)
Temperature dependence of the inelastic intensity at $Q=1.15$ \AA$^{-1}$ and $E=11$ (blue filled circles) and 13 (red filled circles) meV at $E_{f}=30.5$ meV. The lines in the figure are guide to the eye.
}
\end{figure}

As shown in Fig. 2, the spectral weight at low energy is transferred across $\Delta_c$ to the resonance peak. The magnetic spectral weight at $E_{res}$ is enhanced by a factor of two between 35 K and 4 K as shown in Fig. 1.  The doubling of intensity at $Q_{\rm AF}$ is similar to the resonant enhancement observed at the peak energy of 8.6~meV in BaFe$_{1.84}$Co$_{0.16}$As$_{2}$~\cite{Lumsden_09}.  This result is also consistent with calculations assuming $s_{\pm}$ pairing~\cite{Korshunov_08}. A similar intensity is lost at low energy (Fig. 2 (c)). In other words, the dynamical structure factor $S(Q_{AF},E)$ at low energy and low temperature is consistent with the background. The flat energy dependence in the low energy in Fig. 2 (c) may correspond to the full gap state as reported in the penetration depth measurement on a superconducting crystal of PrFeAsO$_{1-\delta}$~\cite{Hashimoto_09}.

\begin{figure}
\includegraphics[width=2.5in]{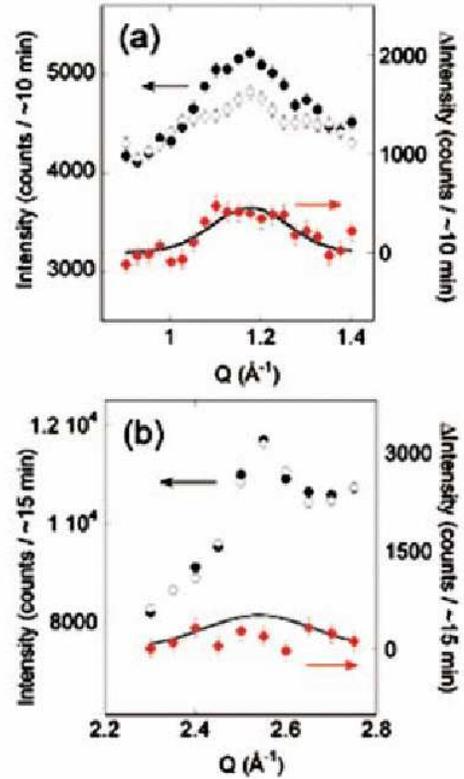}
\caption{(Color online)
(a) Constant-$E$ scans for the first BZ magnetic rods at $T$=4 (black filled circle) and 35 K (open circle) as a function of $Q$ for $E=13$ meV. The $4-35$ K difference for constant-$Q$ scans (red diamond). Solid line is a Gaussian fit. (b) Constant-$E$ scans for the second BZ magnetic rods at $T=4$ (black filled circle) and 35 K (open circle) as a function of $Q$ for $E=13$ meV. The $4-35$ K difference for constant-$Q$ scans (red diamond). Solid line is an expected line based on the fitting in (a) scaled by the square of the Fe$^{2+}$ magnetic form factor.}
\end{figure}

The intensity at the expected resonance position in two different zones, corresponding to (1/2 1/2 $L$) and (3/2 1/2 $L$), are shown in Fig. 4. The solid black line in Fig. 4 (b) is the expected intensity in the second BZ based on scaling the fitted peak at $Q_{\rm AF}=1.15$~\AA$^{-1}$ assuming a purely magnetic excitation.  The intensity is estimated based on Fe$^{2+}$ spherical magnetic form factor and isotropic spin fluctuation ($Q_{\rm AF}=1.15$~\AA$^{-1}$ and 2.55~\AA$^{-1}$).  On the other hand, if there is any phonon contribution in the mode, the peak in the second BZ may become higher than that expected in the first BZ because the phonon intensity increases with increasing $Q$.  This point is particularly important since the resonant mode energy of about $13$ meV coincides with a peak in the phonon density of states. Because of the Fermi surface nesting condition, we may expect phonon anomaly at $Q_{\rm AF}$ across the $T_c$,  as a gap mode observed in an anisotropic s-wave superconductor, YNi$_{2}$B$_{2}$C \cite{Kawano_96,Weber_08}. The resonance mode in the second BZ does not exceed the expectation for magnetic scattering below $T_c$.  This result shows that there may be no phonon contribution in the resonance mode. Strong coupling between magnetism and phonons has been discussed in iron based superconductors~\cite{Yin_08}. However, the observed resonance in this LaFeAsO$_{1-x}$F$_{x}$ system seems to originate predominately from magnetism.  This result is consistent with the $L$-dependence of resonance mode in BaFe$_{1.84}$Co$_{0.16}$As$_{2}$~\cite{Lumsden_09}, where the mode intensity is well described by the Fe$^{2+}$ magnetic form factor.

\begin{figure}
\includegraphics[width=3in]{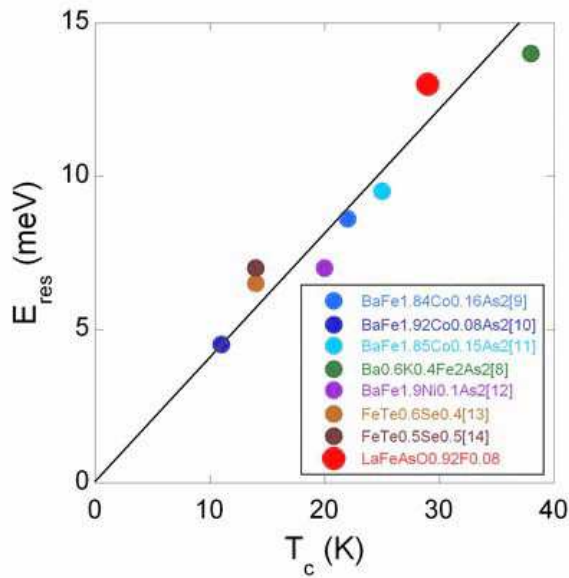}
\caption{(Color online)
Resonance energies $E_{res}$ of iron-based superconductors obtained from neutron inelastic scattering experiments as a function of superconducting transition temperature $T_c$. Solid line is the average slope of 4.7~$k_{B}T_{c}$.}
\end{figure}

The resonant peak energy of $E_{res}\sim13$~meV corresponds to $5.2$~$k_{B}T_{c}$. This energy ratio to $T_c$ is comparable with the ratios observed in other iron based superconductors as shown in Fig. 5. The average peak energy is $4.6(2)$~$k_{B}T_{c}$, suggesting a correlation between the superconductivity and the spin excitations.

Two of the most prominent superconducting pairing symmetries proposed, $s_{\pm}$ \cite{Maier_08,Korshunov_08} and $s_{++}$ \cite{Onari_09}, predict an enhancement in a magnetic excitation spectrum below $T_c$. The predicted peak energies, $E_{res}$, of $s_{\pm}$ and $s_{++}$ are $\sim$1.4 $\Delta_0$ and $>$2 $\Delta_0$, respectively. Based on the observed ratio of 1.38(8) in Ba(Fe$_{1-x}$Co$_{x}$)$_{2}$As$_{2}$~\cite{Lumsden_09}, the present good correlation suggests that $s_{\pm}$ paring is the best to explain the experimental results, although the gap energies are to be measured directly for a definitive statement.


In summary, we have presented the first direct observation of the spin resonance in a member of the $Ln$FeAsO$_{1-x}$F$_x$.  The data are consistent with the resonant intensity being of predominately magnetic origin and the observed resonance energy of 13 meV scales as 5.2 $k_BT_C$ consistent with that seen in other Fe-based superconductors.

We acknowledge F. Esaka, H. Eisaki, and J. A. Fernandez-Baca for their help and fruitful discussions with K. Kakurai, M. Machida, T. Egami, and K. Kuroki.
The experiment was conducted under US-Japan collaboration program and with support of the Grant-in-Aid for Specially Promoted Research, 17001001 and JST, TRIP.  This work was supported by the Scientific User Facilities Division, Office of Basic Energy Sciences, US Department of Energy.



\end{document}